\documentclass[aps,prl,twocolumn,amsmath,amssymb,floatfix, superscriptaddress, 10pt]{revtex4-2}
\usepackage{tabularx}
\usepackage{bm}
\usepackage{euscript}
\usepackage{epsfig,psfrag,subfigure}
\usepackage{graphicx}
\usepackage{color}
\usepackage{amsfonts}
\usepackage{exscale}
\usepackage{amsbsy}
\usepackage{multirow}
\usepackage[normalem]{ulem}
\usepackage[colorlinks=true,linkcolor=blue,citecolor=blue, urlcolor=blue,anchorcolor=blue]{hyperref}
\usepackage{tikz}
\usepackage{xcolor}
\usepackage{tikz}
\usepackage{physics}
\usepackage{amssymb}
\usepackage{mathtools}
\usepackage{amsmath}
\usepackage{dsfont}

\begin{document}
\title{Quantum phase diagram of the extended spin-3/2 Kitaev-Heisenberg model: A DMRG study}
\date{\today}

\author{Gui-Xin Liu}
\thanks{These authors contributed equally.}
\affiliation{School of Physical Science and Technology, ShanghaiTech University, Shanghai 201210, China}

\author{Ting-Long Wang}
\thanks{These authors contributed equally.}
\affiliation{School of Physical Science and Technology, ShanghaiTech University, Shanghai 201210, China}

\author{Yi-Fan Jiang}
\email{jiangyf2@shanghaitech.edu.cn}
\affiliation{School of Physical Science and Technology, ShanghaiTech University, Shanghai 201210, China}

\begin{abstract}
Recently there has been considerable excitement surrounding the promising realization of high-spin Kitaev material, such as the quasi-2D compound CrI$_3$ and CrGeTe$_3$. However, the stability of quantum spin liquids (QSL) against single ion anisotropy (SIA) in these materials and the global quantum phase diagram of the extended spin-3/2 Kitaev model with finite SIA remain unclear.
In this study, we perform large-scale density matrix renormalization group (DMRG) to explore the quantum phase diagram of the generalized spin-3/2 Kitaev-Heisenberg (K-H) model accompanied with SIA $A_c$. In the $A_c=0$ limit, the spin-3/2 K-H model exhibits a quantum phase diagram similar to that of a spin-1/2 system, including two QSLs around antiferromagnetic and ferromagnetic Kitaev models. For models with finite $A_c$, we map out the quantum phase diagram around two Kitaev points and observe distinct types of in-plane vortex orders developed from these two QSL phases. Interestingly, series of nearly degenerate vortex configurations are discovered in each vortex phases. Using linear spin-wave theory, we demonstrate that these vortex configurations can be understood as a consequence of the quantum correction on a continuous family of degenerate classical states.
\end{abstract}

\maketitle

The search for quantum spin liquids (QSLs) in frustrated quantum magnets has been one of the central topics in condensed matter physics for several decades \cite{savaryQuantumSpin2017, broholmQuantumSpin2020}. As an exactly solvable model, the spin-1/2 Kitaev model on the honeycomb lattice provides an unique framework for exploring the novel properties of QSLs \cite{kitaevAnyonsExactly2006, baskaranExactResults2007}. Its ground state is known to host a gapless spin liquid which can be gapped out into a topological phase with non-Abelian excitations by certain time-reversal symmetry breaking perturbations \cite{jiangPossibleProximity2011, jiangFieldinducedQuantum2019, Zhu2020}. 
A large family of 4$d$ and 5$d$ transition metal compounds with strong spin-orbit couplings (SOC) \cite{jackeliMottInsulators2009}, such as $\alpha$-RuCl$_3 $\cite{plumbRuClSpinorbit2014, searsMagneticOrder2015, johnsonMonoclinicCrystal2015a, leahyAnomalousThermal2017, banerjeeNeutronScattering2017, lampen-kelleyAnisotropicSusceptibilities2018, searsFerromagneticKitaev2020, liIdentificationMagnetic2021}, A$_2$IrO$_3$ (A=Na/Li) \cite{choiSpinWaves2012, rauGenericSpin2014a,chaloupkaKitaevHeisenbergModel2010, chaloupkaHiddenSymmetries2015, jiangPossibleProximity2011, yeDirectEvidence2012, chaloupkaZigzagMagnetic2013, yamajiCluesCriteria2016, singhRelevanceHeisenbergKitaev2012}, have been proposed as potential candidates to realize Kitaev-type interactions. 

Recently, there has been increasing interest in exploring higher-spin extensions of the spin-1/2 Kitaev model \cite{baskaranSpinKitaev2008,janssenHoneycombLatticeHeisenbergKitaev2016, consoliHeisenbergKitaevModel2020,stavropoulosMicroscopicMechanism2019, fukuiGroundstatePhase2022a, Ma2023, Cen2023, Georgiou2024, Liu2024}, motivated by potential experimental realizations of Kitaev material in high-spin systems. 
For instance, recent experiments have revealed that the spin-1 Kitaev physics might be realized in novel compounds such as A$_3$Ni$_2$XO$_6$ (A=Li/Na, X=Bi/Sb) \cite{sanoKitaevHeisenbergHamiltonian2018,Zvereva2015,Samarakoon2021,Bera2022,Shangguan2023}, which have sparked extensive theoretical and numerical investigations into the low-energy physics and related quantum phase diagram of the extended spin-1 Kitaev model \cite{koga2018, dongSpin1KitaevHeisenberg2020, Hickey2020, Lee2020, Zhu2020, Khait2021, Bradley2022, Chen2022, Pohle2023,Chen2023}. 
For the case of spin-3/2 systems, several experimental and theoretical research have proposed that the possible Kitaev-type interaction could be realized in CrI$_3$ and CrBr$_3$ \cite{xuInterplayKitaev2018, oitmaaIncipientWelldeveloped2018, kimMicromagnetometryTwodimensional2019, leeFundamentalSpin2020, stavropoulosMagneticAnisotropy2021}. Recent exact diagonalization and DFT calculation suggests that a Kitaev QSL could potentially exist in strained monolayers of CrSiTe$_3$ and CrGeTe$_3$ \cite{xuPossibleKitaev2020}. However, similar to the microscopic models for the spin-1/2 compounds \cite{winterModelsMaterials2017, liuPseudospinExchange2018}, the non-Kitaev interactions are typically not negligible in Kitaev candidates with higher spin momentum and could drive the system away from the Kitaev QSLs. For example, DMRG calculation does not find evidence of QSL phase in the minimal model of CrSiTe$_3$ given by first-principles even when part of non-Kitaev interactions are tuned to zero \cite{zhouStraininducedPhase2021}. The stability of the QSLs against non-Kitaev interactions such as Heisenberg interaction, off-diagonal interaction, and the single-ion term induced by SOC in $S>1/2$ systems and the corresponding quantum phase diagram remain largely unclear.

\begin{figure*}[t]
    \centering
    \includegraphics[width=1.0\linewidth]{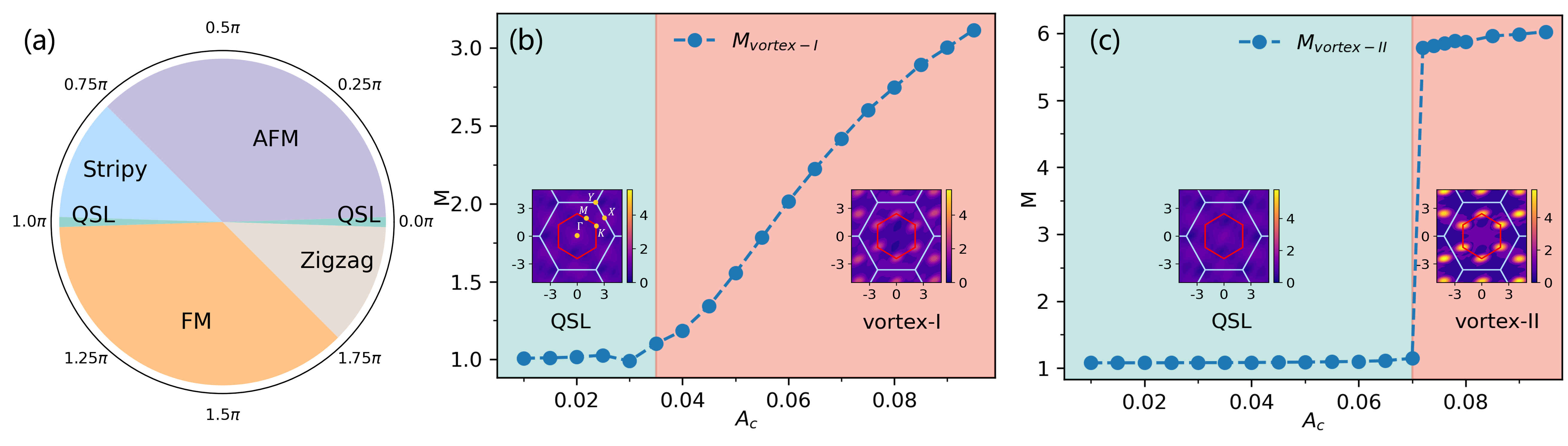}
    \caption{(a) The quantum phase diagram of $S={\frac{3}{2}}$ Kitaev-Heisenberg model without SIA. (b) The phase diagram of the FM Kitaev model with finite $A_c$ measured on $L_y=3$ cylinders. Blue dots represent the order parameter of the vortex order. Inset: the spin structure factor measured at $A_c=0.02$ and $A_c=0.08$ points. (c) The phase diagram of the AFM Kitaev model with finite $A_c$. A tiny FM Heisenberg interaction $J=-3\times10^{-3}$ is introduced to stabilize the vortex-II phase. The vortex order parameter (blue dots) exhibits a clear discontinuity at transition point. Inset: the spin structure factor measured at $A_c=0.01$ and $A_c=0.09$ points. }
    \label{fig:phase_0pi_1pi}
\end{figure*}

In this work, we systematically investigate the quantum phase diagram of the extended spin-3/2 Kitaev-Heisenberg (K-H) model accompanied by the out-of-plane single-ion anisotropy (SIA) on both $L_y=3$ and $4$ cylinders. 
The inclusion of the SIA term significantly changes the quantum phase diagram in the vicinity of the two Kitaev points. Our calculation reveals that both two Kitaev QSLs become fragile under a small SIA. At the ferromagnetic (FM) Kitaev point ($K<0$), SIA induces a continuous phase transition into an in-plane vortex phase characterized by the ordering vector at $K$ point of the Brillouin zone (see Fig.\ref{fig:phase_0pi_1pi}(b)). 
A similar SIA induced vortex phase is observed in the antiferromagnetic (AFM) Kitaev model with a tiny FM Heisenberg interaction ($K>0$, $J\sim 0^-$), but the phase transition is first order.
Interestingly, on the $J\sim 0^+$ side of the same model, we also find a first-order phase transition between the QSL and AFM phases, indicating that a triple point involving the AFM, vortex and QSL phases exists in the AFM Kitaev model with small $A_c$.
The spin patterns of the two coplanar vortex phases are similar to those of the vortex states in the spin-1/2 $J-K-\Gamma-\Gamma'$ model for iridates \cite{chaloupkaHiddenSymmetries2015} and the projected spin configurations of the noncoplanar vortex phases observed in the K-H model with a uniform magnetic field $\mathbf{h}$ $||$ [111] \cite{janssenHoneycombLatticeHeisenbergKitaev2016, consoliHeisenbergKitaevModel2020}. However, the vortex phases in the spin-3/2 model are established without the off-diagonal spin interaction and external magnetic field.  Similar vortex phases have also been observed in the spin-1 K-H model with onsite anisotropy \cite{singhaniaEmergencevortex2023}. 
Though the hidden SU(2) symmetry \cite{chaloupkaHiddenSymmetries2015} is absent in the model with finite SIA, we still discover several nearly degenerate spin configurations connected by the opposite spin rotation of certain angles on two sublattices in each vortex phase. Using classical spin analysis and linear spin-wave theory, we demonstrate that these degenerate spin patterns in the two vortex phases are selected from a continuous family of degenerate classical states via the order-by-disorder mechanism driven by the quantum fluctuations.

{\it Model and Method:}
We use large-scale DMRG method \cite{whiteDensitymatrixAlgorithms1993} investigate the quantum phase diagram of the spin-$3/2$ Kitaev-Heisenberg model accompanied with out-of-plane SIA, 
\begin{equation}  \label{eq:Ham}
    H=\sum_{\langle i,j\rangle _{\gamma}}{\left( KS_{i}^{\gamma}S_{j}^{\gamma}+J\mathbf{S}_i\cdot \mathbf{S}_j \right)}+\frac{A_c}{3}\sum_i{(}\sum_{\alpha}{S_{i}^{\alpha}})^2,
\end{equation}
where $S_i^\alpha$ is the $\alpha=x,y,z$ component of spin $S=3/2$ operator on site $i$ and $\left< i,j \right>_\gamma$ stands for the $\gamma={X,Y,Z}$ bond of the Kitaev model. The nearest neighbor Kitaev interaction and Heisenberg interaction are parameterized as $K=\cos{\theta}$ and $J=\sin{\theta}$, respectively. The second term of the model is the single ion anisotropy originated from the strong SOC in spin-3/2 systems.
In this paper, we focus on ground-state phase diagram of the model with experimentally relevant [111] SIA on $L_y = 3$ and $L_y=4$ cylinders of length up to $L_x = 9$ unit cell. The total number of sites is $N=2 \times L_x \times L_y$. In our DMRG simulation, we keep up to $m=3000$ states in each DMRG block and perform at least 20 sweeps to ensure the convergence of numerical results.

{\it The spin-3/2 Kitaev-Heisenberg model:} 
It has been established from previous parton mean-field study \cite{jinUnveilingKitaev2022} and DMRG calculation \cite{natoriQuantumLiquids2023a} that the spin-3/2 Kitaev model hosts a gapless QSL ground state. After introducing the nearest neighbor Heisenberg interaction, recent functional renormalization group \cite{fukuiGroundstatePhase2022a} and coupled cluster method \cite{Georgiou2024} studies have shown that the quantum phase diagram of the spin-3/2 K-H model exhibits the structure similar to that for the spin-1/2 system \cite{chaloupkaZigzagMagnetic2013}. Here, we use DMRG calculate the phase diagram of the spin-3/2 K-H model on $L_y=3$ cylinder. Consistent with previous results \cite{fukuiGroundstatePhase2022a,Georgiou2024}, we observe the two QSL phases in the vicinity of the FM ($\theta\sim \pi$) and AMF ($\theta\sim 0$) Kitaev points, together with four magnetic ordered phases as shown in Fig.\ref{fig:phase_0pi_1pi}(a). The magnetic properties of the ordered phases are investigated by calculating the spin-spin correlation function and the static spin structure factor (SSSF) defined as 
\begin{equation}
	S^{\alpha \alpha} (\mathbf{k}) = \frac{1}{N} \sum_{i,j} \left\langle S^{\alpha}_i S^{\alpha}_j  \right\rangle e^{i \mathbf{k} (\mathbf{r}_i - \mathbf{r}_j) }  .
\end{equation}
The phase boundaries in Fig.\ref{fig:phase_0pi_1pi}(a) are determined by the second-order derivative of the ground state energy as a function of the parameter $\theta$. Compared to the phase diagrams of the spin-1/2 and spin-1 K-H models \cite{chaloupkaZigzagMagnetic2013, dongSpin1KitaevHeisenberg2020}, the regions of the two QSLs in the spin-3/2 model are narrower (e.g., $-0.001\pi \lesssim \theta \lesssim 0.0015\pi$ for AFM Kitaev cases) due to the reduction of the quantum fluctuation in high-spin systems. In the following, we study the evolution of the two QSL phases and adjacent ordered phases of the K-H model in the presence of SIA.

\begin{figure}[btp]
    \centering  
    \includegraphics[width=1.0\linewidth]{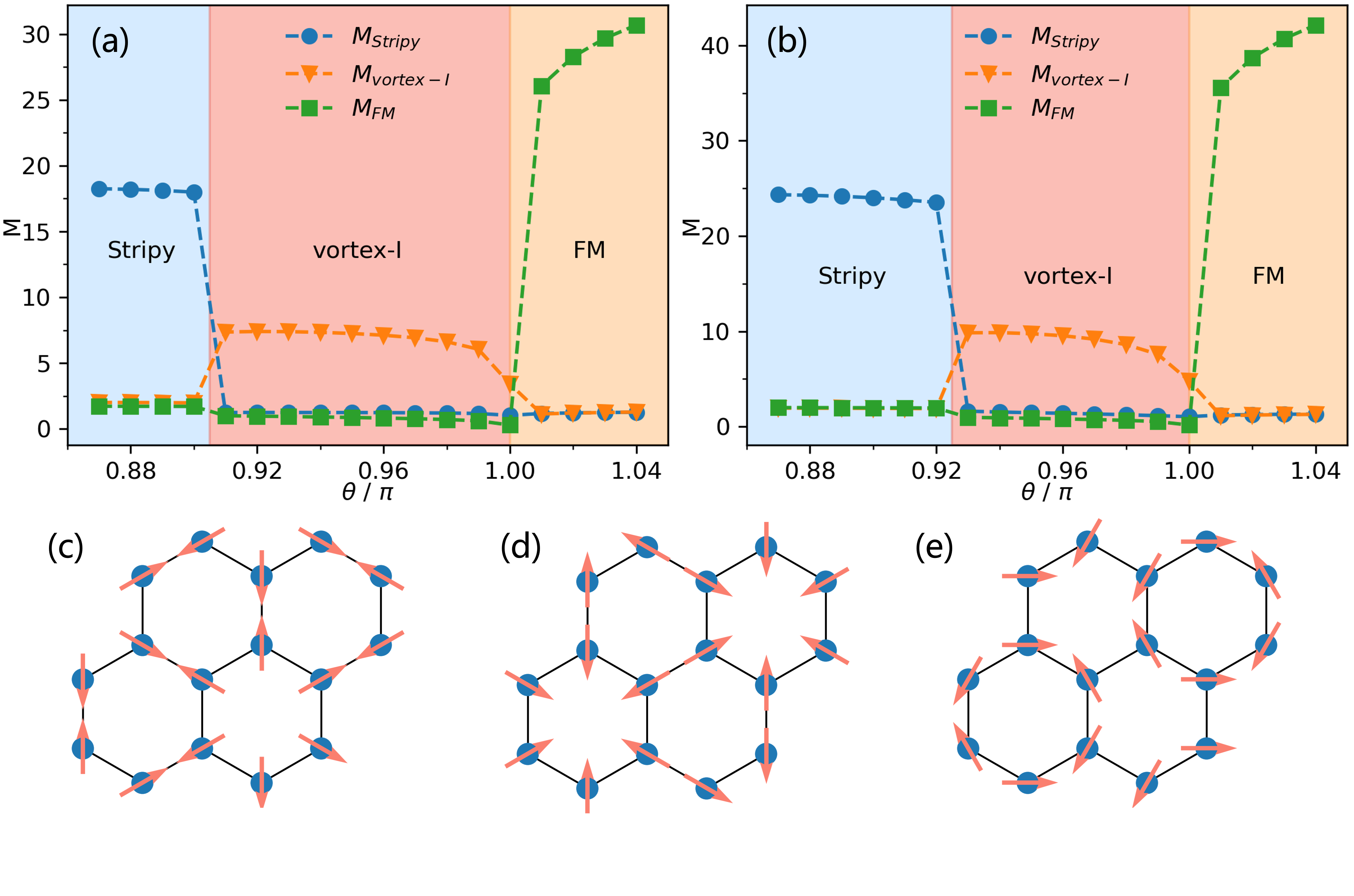}
    \caption{(a) The quantum phase diagram of the $A_c=0.5$ model with $\theta$ ranged from $0.86\pi$ to $1.05\pi$ on $L_y=3$ cylinders. The phase boundaries are determined by the changes of the magnetic order parameters shown in different colors. (b) Phase diagram of the same models on $L_y=4$ cylinders. (c-e) Selected spin configurations in vortex-I phases. These nearly degenerate spin pattern can be connected by the opposite in-plane spin rotation on the two sublattices. }
    \label{fig:phase_theta_pi}
\end{figure}

{\it Effect of SIA:}
We first investigate the evolution of the two QSL phases under increasing SIA. As depicted in the phase diagrams shown in Fig.\ref{fig:phase_0pi_1pi}(b) and (c), both QSL phases are stable until the SIA strength exceeds certain thresholds.
For the $\theta=\pi$ case, we observe the emergence of a long-range magnetic ordered phase on three-leg cylinders when $A_c > 0.035$. However, the ordering momentum of this phase does not correspond to any of the four ordered phases previously identified in the K-H model. At $A_c=0.08$ point deep inside the ordered phase, the spin structure factor $S^{\alpha \alpha} (\mathbf{k})$ shown in the inset of Fig.\ref{fig:phase_0pi_1pi}(b) exhibits sharp peaks at the $\mathbf{K}$ point of the Brillouin zone. The corresponding spin pattern, referred to as the vortex-I phase, is illustrated in Fig.~\ref{fig:phase_theta_pi}(e), where the spins are coplaner and form vortex-like structures on hexagons. This vortex spin structure closely resembles the vortex phase generated by the $\mathcal{T}_6$ transformation in the spin-1/2 K-$\Gamma$ model with equivalent Kitaev and off-diagonal interactions \cite{chaloupkaHiddenSymmetries2015}. However, in the spin-3/2 model, the vortex phases are induced by onsite anisotropy rather than a strong off-diagonal $\Gamma$ interaction.
The nature of the quantum phase transition observed in Fig.\ref{fig:phase_0pi_1pi}(b) is determined by measuring the order parameter $M= \sqrt{S^{\alpha \alpha} (\mathbf{K})}$ of the vortex order. As $A_c$ increases across the critical point $A_c\sim 0.035$, $M$ does not exhibit any sharp discontinuity, indicating a second-order phase transition from QSL to the SIA-induced vortex-I phase.

\begin{figure}[bt]
\centering
\includegraphics[width=1.0\linewidth]{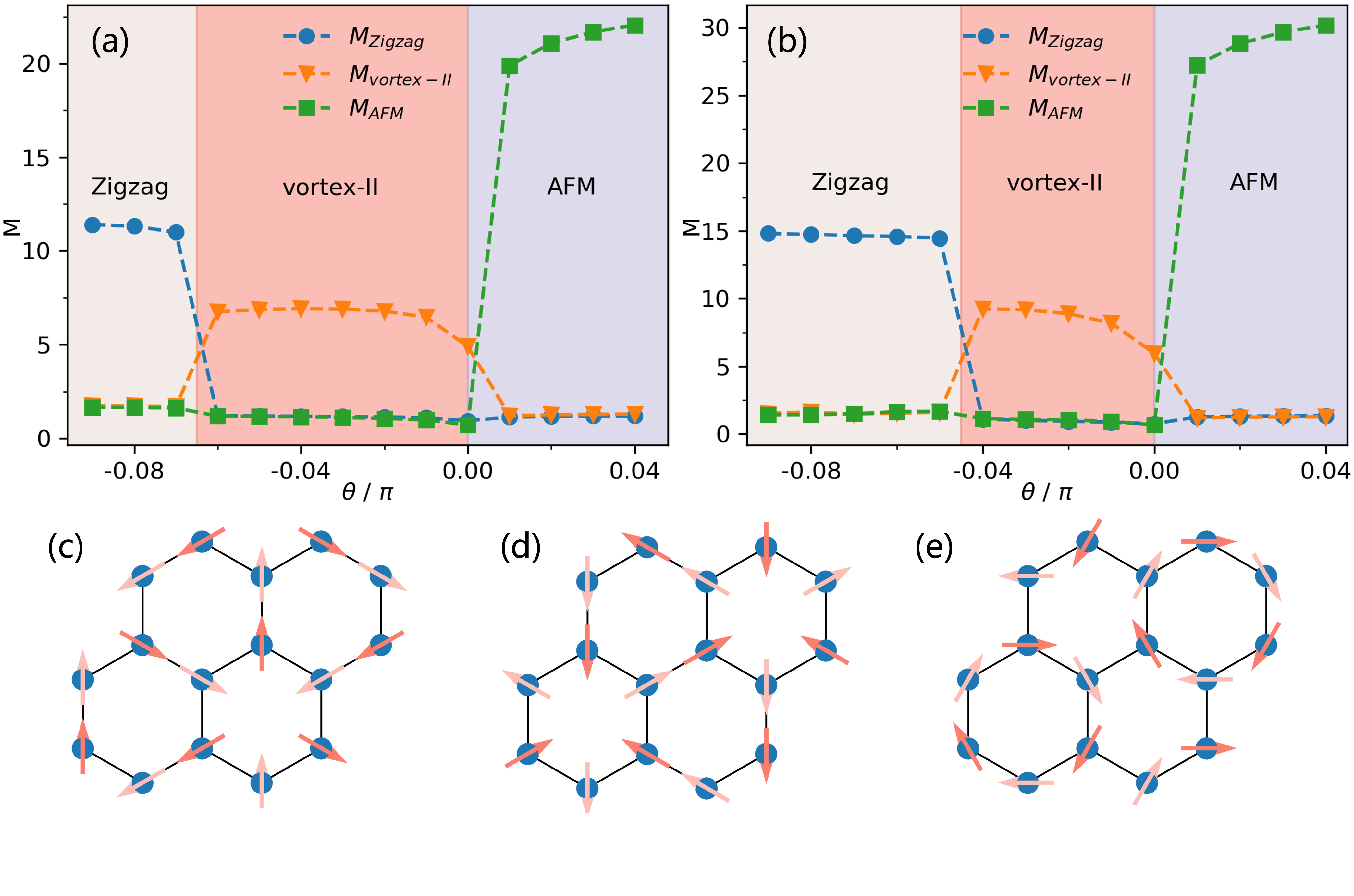}
\caption{(a) The quantum phase diagram of the $A_c=0.5$ model with $\theta$ ranged from $-0.1\pi$ to $0.05\pi$ on $L_y=3$ cylinders. (b) Phase diagram of the same models on $L_y=4$ cylinders. The order parameters at the $\theta=0$ point represent the results obtained from the $\theta=0^-$ (i.e. $K>0, J=0^-$) side of the transition point. (c-e) Selected spin configurations in vortex-II phases. }
\label{fig:phase_theta_0}
\end{figure}

A similar SIA-induced vortex phase, the vortex-II phase, is obtained from the QSL phase in the AFM Kitaev model with $K>0$. 
The spin structure factor obtained in the vortex-II phase exhibits the same momentum-$\mathbf{K}$ peaks as in the vortex-I phase, but their real-space spin pattern differs by a global $\pi$-phase spin rotation on one sublattice, as illustrated in Fig.\ref{fig:phase_theta_0}(c). 
The transition behavior from QSL to the vortex-II phase is also different from that in the vortex-I case. Following the same procedure, we identify that a sharp discontinuity of the vortex order parameter occurs at a relatively larger critical point $A_c \sim 0.07$ (see Fig.\ref{fig:phase_0pi_1pi}(c)), implying a first-order phase transition between the two phases.
Interestingly, our calculation shows that a tiny FM Heisenberg interaction, e.g. $J/K \sim -3\times10^{-3}$, is necessary to stabilize the vortex-II phase. Once the Heisenberg $J$ becomes positive, the entire vortex-II phase in the large SIA region is replaced by an AFM phase characterized by sharp peaks located at $\Gamma'$ point. The phase transition between the QSL and AFM phase is also first order.
This result suggests that the $K=1$ and $A_c\sim 0.07$ point happens to be a triple point of three first-order transition lines connecting the QSL, AFM, and vortex-II phases.

{\it Quantum phase diagram with finite SIA:}
To comprehensively understand the rich SIA-induced phases emerging from QSLs, we examine the ground-state properties of the K-H model in the presence of finite SIA near the two Kitaev points. Starting from the phase diagram of the K-H model illustrated in Fig.\ref{fig:phase_0pi_1pi}(a), we consider a relatively large SIA $A_c=0.5$ and explore the $A_c$ induced phases around the $\theta=\pi$ and $\theta=0$ points on $L_y=3$ and $L_y=4$ cylinders. 

The SIA-induced phase diagram around the FM Kitaev point ($\theta \sim \pi$) is depicted in Fig.\ref{fig:phase_theta_pi}(a) and (b). On the $L_y=3$ cylinder, the vortex-I phase dominates the $\theta \in [0.91\pi, \pi]$ region of the phase diagram, sandwiched by the stripy phase characterized by the peak of SSSF at $\mathbf{X}$ point and the FM phase with the peak of SSSF at $\mathbf{\Gamma'}$ point. The first-order phase transition boundaries between these phases are identified by the abrupt changes in the corresponding order parameters illustrated in Fig.\ref{fig:phase_theta_pi}(a) and (b). Intriguingly, in the vortex-I phase, we observe several nearly degenerate vortex patterns that can be connected by the opposite spin in-plane rotations on two sublattices of the honeycomb lattice, with discretized rotation angle $\alpha$. 
The configurations related by $\alpha=\pi/3$ rotation (e.g. Fig.\ref{fig:phase_theta_pi}(c) and (d)) shared the same energy, while the energies of those related by $\alpha=\pi/6$ rotation (Fig.\ref{fig:phase_theta_pi} (e)) are slightly different. In most part of the vortex-I phase, i.e. $\theta \lesssim 0.97\pi$, the configuration illustrated in Fig.\ref{fig:phase_theta_pi}(c) and (d) possesses the lowest energy. However, close to the Kitaev point, the energies of the patterns in Fig.\ref{fig:phase_theta_pi}(c) and (e) become nearly identical. 

Around the AFM Kitaev point, finite onsite anisotropy induces a phase diagram strikingly similar to that near the FM Kitaev interaction. 
As shown in Fig.~\ref{fig:phase_theta_0}(a) and (b), the phase diagram for $A_c=0.5$ involves zigzag, vortex-II, and AFM phases as $\theta$ varies from -0.1$\pi$ to 0.05$\pi$. Like the vortex-I phase that terminates at the FM Kitaev point, the vortex-II phase in the center of the phase diagram ends right at the AFM Kitaev point $\theta=0$. In the vortex-II phase, we also observe a set of nearly degenerate ground states with different in-plane vortex configurations, as illustrated in Fig.\ref{fig:phase_theta_0}(c-e). These spin configurations in the vortex-II phase have a one-to-one correspondence to those in the vortex-I phase: a global $\pi$-phase rotation along the [111] axis on one sublattice can transform one vortex configuration into the other. Furthermore, the energy dependence of the patterns in the vortex-II phase resembles that in the vortex-I phase. For instance, the patterns in Fig.\ref{fig:phase_theta_0}(c) and (d) possess the same lowest energy in the region $\theta \lesssim 1.98\pi$ in the vortex-II phase.

We also explore the potential finite-size effect by investigating the ground-state properties of the same models on wider $L_y=4$ cylinders. As presented in Fig.\ref{fig:phase_theta_pi}(b) and Fig.\ref{fig:phase_theta_0}(b), we find that all phases persist on the wider cylinder as their order parameters increase roughly linearly with $L_y$. However, the two vortex phases shrink on the $L_y=4$ cylinder, which could be attributed to frustration of the vortex pattern in the $L_y=4$ system.

Given the similarities between the vortex-I and vortex-II phases, it is natural to ask whether these phases can be understood in a unified way. In the following, we demonstrate that the degenerate ground states in both vortex phases can be explained by the order-by-disorder mechanism arising from quantum corrections on the continuous sets of the degenerate ground states of the classical spin models \cite{priceFinitetemperaturePhase2013,janssenHoneycombLatticeHeisenbergKitaev2016, consoliHeisenbergKitaevModel2020}.

\begin{figure}[bt]
    \centering
    \begin{subfigure}{}
        \includegraphics[scale=0.32]{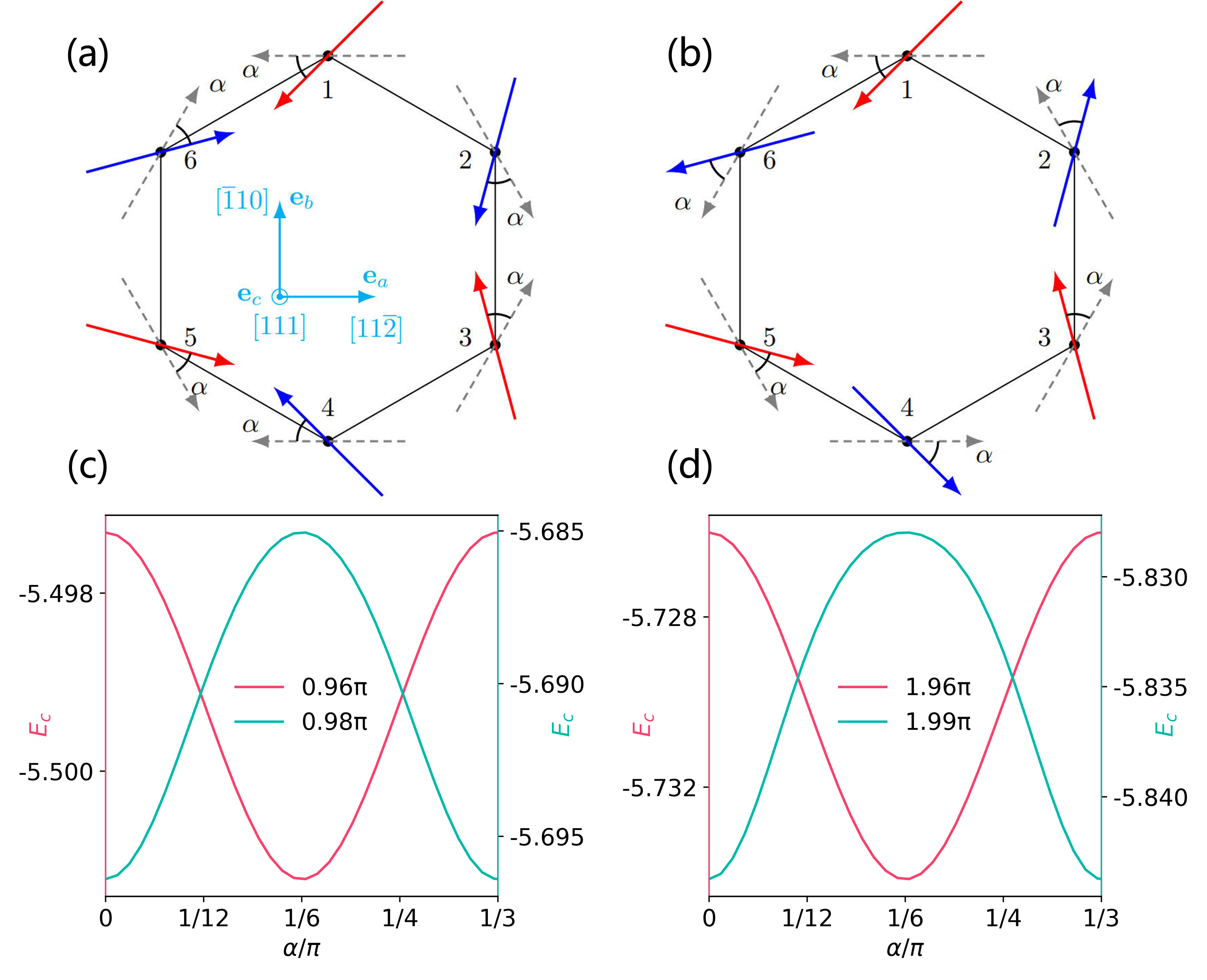}
    \end{subfigure}
    \caption{(a) Illustration of classical vortex patterns in vortex-I phase. Dashed arrows represent the vortex configuration observed in DMRG calculation. The degenerate spin pattern are constructed by opposite spin rotations on two sublattices labeled in different colors.
    (b) Classical spin patterns in vortex-II phase defined in the similar way.
    (c) Corrected energies of the unit cell obtained by the linear spin-wave theory based on the rotated pattern for the vortex-I phase. In the models close to the Kitaev point ($\theta=0.98\pi$), the minimal of the corrected energy is at $\alpha=0$. For the $\theta=0.96\pi$ case, the minimal located at $\alpha=\pi/6$. (d) Corrected energy of the unit cell of the classical states in the vortex-II phase as a function of $\alpha$. The minimal of the corrected energies also happens at $\alpha=0$ or $\pi/6$ depending on the value of $\theta$.
    }
    \label{fig:vortex_lsw}
\end{figure}

{\it Order-by-disorder in vortex phases:} As a first step to understand the nature of the vortex phases, we analyze the ground-state structures of the K-H model with large SIA in the classical limit. 
Here we focus on the large $A_c$ region where spins lie in the [111] plane. In this classical model, all spin patterns observed in vortex-I phases (see Fig.~\ref{fig:phase_theta_pi}(c-e)) are degenerate, with energy per site $E_{\text{vor-I},0}=\cos\theta S^2/2$. This energy is lower than that of the stripy phase $E_{\text{str},0}=(\cos(\theta)/6 -\sin (\theta)/2)S^2$ and FM phase $E_{\text{FM},0}=(3\sin\theta+\cos\theta)S^2/2$ over a broad range $\theta \in (0.813\pi,\pi)$. Moreover, as shown in Fig.~\ref{fig:vortex_lsw}(a), starting from the spin patterns obtained from DMRG calculation in the vortex-I phases (indicated by dashed arrows), an arbitrary rotation of spins on the A and B sublattice in opposite ways along the [111] axis leaves the energy of the classical states unchanged, which leads to an infinitely degenerate ground-state manifold characterized by the U(1) spin rotation angle $\alpha$. Similar classical degeneracies are also observed in the models for the vortex-II phase, as shown in Fig.~\ref{fig:vortex_lsw}(b).

These extensive classical degeneracies can be lifted through an order-by-disorder mechanism driven by quantum fluctuations \cite{priceFinitetemperaturePhase2013,janssenHoneycombLatticeHeisenbergKitaev2016, consoliHeisenbergKitaevModel2020}. To incorporate the leading order of quantum corrections, e.g. zero-point quantum fluctuation, we 
utilize linear spin-wave theory based on rotated spin configurations characterized by the rotation angle $\alpha$ shown in Fig.~\ref{fig:vortex_lsw}. 
In the continuum limit, the ground-state energy including quantum fluctuations up to the order of $O(S)$ can be expressed as an integral of the magnetic bands over the first Brillouin zone:
\begin{equation}
E=-\frac{S(S+1)}{2} |\cos\theta|+\frac{V}{24\pi^2}
\sum_{\gamma=1}^6 \int_{\text{BZ}} \mathrm{d}^2\mathbf{k}\ 
\omega_\gamma(\mathbf{k})
\end{equation}
where $V=9\sqrt{3}a^2/2$ is the area of the primitive unit cell, $a$ is the lattice constant, and $\omega_\gamma(\mathbf{k})$ represents the magnetic band corresponding to the classical configuration. The detailed definitions of $\omega_\gamma(\mathbf{k})$ are provided in the supplemental materials (SM).
The quantum corrections of the energies for the models with four selected $\theta$ are shown in Fig.\ref{fig:vortex_lsw} (c) and (d). 
In both vortex-I and vortex-II cases, it can be proved that the spin-wave spectra are invariant under $\alpha\rightarrow -\alpha$ and $\alpha\rightarrow \alpha+\pi/3$ (see SM for details), which results in $E(\alpha)=E(-\alpha)=E(\alpha+\pi/3)$. This relation explains the degeneracy of the vortex pattern observed in DMRG simulations, such as the two degenerate configurations in Fig.~\ref{fig:phase_theta_pi}(c) and (d).
The calculation of the quantum corrections further reveals that the minimum of the corrected energy $E(\alpha)$ occurs at $\alpha_{\text{min}}=0$ or $\pi/6$. For models close to Kitaev points ($\theta=0$ or $\pi$), the minimum of energy is achieved at $\alpha=0$, corresponding to the spin patterns shown in Fig.~\ref{fig:phase_theta_pi}(e) and Fig.~\ref{fig:phase_theta_0}(e). When $\theta$ is tuned slightly away from $0$ or $\pi$ points, the minimum moves to $\alpha=\pi/6$, providing patterns consistent with those shown in Fig.~\ref{fig:phase_theta_pi}(c) and Fig.~\ref{fig:phase_theta_0}(c). 
Moreover, our analysis of the spectrum indicates that the vortex-I phase is stable only in the region $\theta\in(0.87\pi, \pi)$. While for the vortex II phase, the stable region is $\theta \in ( 1.87\pi, 2\pi)$. 
These results are consistent with the DMRG phase diagram depicted in Fig.\ref{fig:phase_theta_pi} and \ref{fig:phase_theta_0}, where both vortex phases terminate at the two Kitaev points.

{\it Conclusion:} In this study, we systematically explore the quantum phase diagram of the S=3/2 Kitaev-Heisenberg model with single ion anisotropy. Our DMRG results reveal that the two spin liquid phases in the AFM and FM Kitaev models are highly sensitive to weak SIA, whereas the conventional magnetic ordered phases appearing in the K-H model remain stable even at larger SIA ($A_c\sim 0.5$). The increasing SIA triggers two distinct transitions from the QSLs: a continuous phase transition to the vortex-I phase around the FM Kitaev point and a first-order phase transition to the vortex-II phase around the AFM Kitaev point. In each vortex phase, we observe a set of nearly degenerate low-energy states with different vortex configurations. 
Using linear spin-wave theory, we demonstrate that all of the vortex states observed in the DMRG calculation can be understood as low-energy states selected by the quantum fluctuations on a continuous set of degenerate classical spin states. Interestingly, part of these degenerate spin patterns (or their projected version) also appear in certain regions of the phase diagrams of extended Kitaev models, such as the K-$\Gamma$ models \cite{chaloupkaHiddenSymmetries2015} and K-H models in the presence of an external magnetic field \cite{janssenHoneycombLatticeHeisenbergKitaev2016, consoliHeisenbergKitaevModel2020}. Exploring the potential relationships between these seemingly distinct models could provide new insight into the intricate phase diagrams of the high-spin Kitaev systems. It will also be interesting to study the high-spin Kitaev models that include longer-range and off-diagonal couplings in the future work, as such extensions may lead to the discovery of more complex SIA-induced phases.
 
{\it Acknowledgement:} This work is supported in part by the National Key R$\&$D Program of China under Grants No. 2022YFA1402703, NSFC under Grant No. 12347107, and Shanghai Pujiang Program under Grant No.21PJ1410300.

\bibliography{draft}

\clearpage
\newpage
\appendix
\onecolumngrid
\begin{center}
\textbf{\large Supplemental Material for ``Quantum phase diagram of the extended spin-3/2 Kitaev Heisenberg model on three- and four-leg cylinders''}
\end{center}
\twocolumngrid

\setcounter{figure}{0}
\renewcommand{\thefigure}{S\arabic{figure}}
\renewcommand{\theHfigure}{S\arabic{figure}}
\setcounter{equation}{0}
\renewcommand{\theequation}{S\arabic{equation}}
\setcounter{section}{0}
\renewcommand{\thesection}{\Roman{section}}
\setcounter{secnumdepth}{4}

\section{Linear spin-wave theory for vortex states}

\subsection{Linear spin-wave formalism}
Here we present the detail of the linear spin-wave analysis for the vortex phase. The direction of the classical in-plain spin is parameterized by the angle $\phi$ relative to the basis vector $\mathbf{e}_a=(1,1,-2)/\sqrt{6}$.
According to the DMRG results and Fig. \ref{fig:vortex_lsw} in the main text, the classical vortex patterns are defined by 
\begin{equation} \label{vortexdef}
\begin{aligned}
\phi_{\gamma} &= 
(-1)^{\gamma-1}\alpha+\pi+(\gamma-1)\Delta\\
\Delta&=\begin{cases}
2\pi/3,& \text{vortex I }(\theta \approx \pi)  \\
-\pi/3,& \text{vortex II }(\theta \approx 0)  ,
\end{cases}
\end{aligned}
\end{equation}
where $\phi_\gamma$ measures the orientation of the classical spin on sublattice $\gamma$ ($\gamma=1,2,\cdots,6$) relative to the basis vector $\mathbf{e}_a$. The Holstein-Primakoff transformation is then applied to spin operators on each site:
\begin{equation}
\begin{cases}
S^+_i=\left(2S-a^\dagger_i a_i\right)^{1/2} a_i \\
S^-_i=a^\dagger_i \left(2S-a^\dagger_i a_i\right)^{1/2} , \\
S^3_i=S-a^\dagger_i a_i
\end{cases}
\end{equation}
where $S^3$ is the direction of classical spin. In first order approximation, they become
\begin{equation}
\begin{cases}
S^1_i\approx \sqrt{S/2}\ (a_i+a^\dagger_i) \\
S^2_i\approx -i\sqrt{S/2}\ (a_i-a^\dagger_i) , \\
S^3_i=S-a^\dagger_i a_i
\end{cases}
\end{equation}
where $S^2$ is set to along the $[111]$ direction. In this basis, the spin operators on site $i$ can be presented as
\begin{equation} \label{abccomponents}
\begin{aligned}
\mathbf{S}_i(\phi_{\gamma_i})&=\left(S^3_i\cos\phi_{\gamma_i}
-S^1_i\sin \phi_{\gamma_i}\right)\mathbf{e}_a\\ 
&\quad+\left(S^3_i\sin\phi_{\gamma_i}+S^1_i
\cos \phi_{\gamma_i}\right)\mathbf{e}_b \\
&\quad+S^2_i\ \mathbf{e}_c , 
\end{aligned}
\end{equation}
where $\mathbf{e}_b=(-1,1,0)/\sqrt{2}$ and $\mathbf{e}_c=(1,1,1)/\sqrt{3}$. Rewriting the Hamiltonian in boson operators and keeping up to $O(S)$ order gives
\begin{equation}
\begin{aligned}
H(\alpha,\theta,A_c)&=H_{\text{classical}}(\theta)
+\sum_i H_{1}(i;\alpha,\theta) a_i^\dagger a_i \\
&+\sum_{\left<ij\right>}
\begin{pmatrix}
a_i^\dagger& a_i
\end{pmatrix}
\mathbf{H}_{2}(i,j;\alpha,\theta)
\begin{pmatrix}
a_j \\a_j^\dagger 
\end{pmatrix}\\
&-\frac{S A_c}{2}\sum_i \left(a_i-a_i^\dagger\right)^2 .
\end{aligned}
\end{equation}
Detailed calculation shows that $H_{\text{classical}}$ is independent of $\alpha$, 
and no linear terms in $a$ or $a^\dagger$ are present. 
We perform Fourier transformation of real-space operators:
\begin{equation}
a_i=\frac1{\sqrt{N}}\sum_{\mathbf{k}} e^{i\mathbf{k}\cdot \mathbf{R}_i}
a_{\mathbf{k},\gamma},\ 
a_i^\dagger=\frac1{\sqrt{N}}\sum_{\mathbf{k}} e^{-i\mathbf{k}\cdot \mathbf{R}_i}
a^\dagger_{\mathbf{k},\gamma} . 
\end{equation}
The quadratic part of the Hamiltonian can be written in the following form:
\begin{equation}
\sum_{\mathbf{k}}
\begin{pmatrix}
a_{\mathbf{k},\gamma}^\dagger & a_{-\mathbf{k},\gamma}
\end{pmatrix}
\mathbb{H}(\mathbf{k};\alpha,\theta,A_c)
\begin{pmatrix}
a_{\mathbf{k},\gamma} \\ a_{-\mathbf{k},\gamma}^\dagger
\end{pmatrix} , 
\end{equation}
where $\mathbb{H}$ is a $12\times 12$ Hermitian matrix with form
\begin{equation} \label{hamform}
\mathbb{H}(\mathbf{k})=
\begin{pmatrix}
\mathbf{A}(\mathbf{k}) & 
\mathbf{B}(\mathbf{k})\\
\mathbf{B}^*(-\mathbf{k}) & 
\mathbf{A}^*(-\mathbf{k})
\end{pmatrix} .
\end{equation}
We employ multi-flavour bose Bogoliubov transformation to solve eigenmodes of the system:
\begin{eqnarray}
&&\begin{pmatrix}
a_{\mathbf{k},\gamma} \\ a_{-\mathbf{k},\gamma}^\dagger
\end{pmatrix}
=\mathbf{U}(\mathbf{k})
\begin{pmatrix}
b_{\mathbf{k},\gamma} \\ b_{-\mathbf{k},\gamma}^\dagger
\end{pmatrix} , \\
&&\mathbf{U}(\mathbf{k})\sigma_z \mathbf{U}^\dagger(\mathbf{k})=\sigma_z . \nonumber
\end{eqnarray}
Under the basis of $b_{\mathbf{k},\gamma}$ and 
$b_{-\mathbf{k},\gamma}^\dagger$, the matrix of the Hamiltonian 
becomes diagonal:
\begin{equation}
\text{diag}(\omega_\gamma(\mathbf{k}),\omega_\gamma(-\mathbf{k}))
:=\mathbb{D}(\mathbf{k})=
\mathbf{U}^\dagger(\mathbf{k})\mathbb{H}(\mathbf{k})
\mathbf{U}(\mathbf{k})
\end{equation}
It can be proved that the spectrum $\omega_\gamma$ is 
well-defined as a function of $\mathbf{k}$, though the band ordering 
in $\gamma$ is interchangeable. 

\subsection{Corrections to classical energy}
The full Hamiltonian of the linear spin-wave system is
\begin{equation}
\begin{aligned}
&H(\alpha,\theta,A_c)=H_{\text{classical}}(\theta)
-\frac12 \sum_i H_{1}(i;\alpha,\theta)\\
&\quad+\sum_{\mathbf{k}}
\begin{pmatrix}
a_{\mathbf{k},\gamma}^\dagger & a_{-\mathbf{k},\gamma}
\end{pmatrix}
\mathbb{H}(\mathbf{k};\alpha,\theta,A_c)
\begin{pmatrix}
a_{\mathbf{k},\gamma} \\ a_{-\mathbf{k},\gamma}^\dagger
\end{pmatrix}
\end{aligned}
\end{equation}
The 2nd term of RHS comes from exchanging $a_{i}^\dagger$ and 
$a_{i}$. There's a relation between this term and the classical energy:
\begin{equation}
\sum_i H_{1}(i;\alpha,\theta)=-2H_{\text{classical}}(\theta)/S
\end{equation}
which is due to changing $S_i^3$
from $S$ to $S-a_i^\dagger a_i$ in Holstein-Primakoff transformation. 
Since $H_{\text{classical}}\sim S^2$, adding the 2nd term to the classical 
energy is equivalent to 
replacing $S^2$ with its quantum version $S(S+1)$. The Hamiltonian in terms 
of eigenmodes $b_{\mathbf{k},\gamma}$ and $b_{-\mathbf{k},\gamma}^\dagger$ 
is:
\begin{equation}
\begin{aligned}
&H(\alpha,\theta,A_c)=-3N|\cos\theta|S(S+1)\\
&\quad+\sum_{\mathbf{k}}\sum_{\gamma=1}^6 
\omega_\gamma(\mathbf{k};\alpha,\theta,A_c)
\left( 2b_{\mathbf{k},\gamma}^\dagger b_{\mathbf{k},\gamma}+1 \right)
\end{aligned}
\end{equation}
where $N$ is the total number of primitive cells. In the continuum limit, the per-site ground state energy up to the leading order of quantum correction is
\begin{equation}
E=-\frac{S(S+1)}{2} |\cos\theta|+\frac16\frac{V}{(2\pi)^2}
\sum_{\gamma=1}^6 \int_{\text{BZ}} \mathrm{d}^2\mathbf{k}\ 
\omega_\gamma(\mathbf{k}) ,
\end{equation}
where $V=9\sqrt{3}a^2/2$ is the area of primitive unit cell, $a$ is the lattice space. The spectrum $\omega_\gamma(\mathbf{k})$ can be simply obtained by numerical diagonalization of $\sigma_z \mathbb{H}(\mathbf{k})$.

\subsection{Invariant spectrum under $\alpha\rightarrow \pm \alpha +\pi/3$}
We show that casting $\alpha$ to $\alpha+\pi/3$ preserves the spectrum of the linear spin-wave model. By definition \ref{vortexdef} of rotation angle $\phi_\gamma$, we find 
\begin{equation}
\phi_\gamma(\alpha+\pi/3)=\phi_{\gamma'}(\alpha)+\pi ,
\end{equation}
where $\gamma'=[\gamma-2(-1)^{\gamma}]$ mod $6$. 

\begin{figure}
    \centering
    \includegraphics[width=0.7\linewidth]{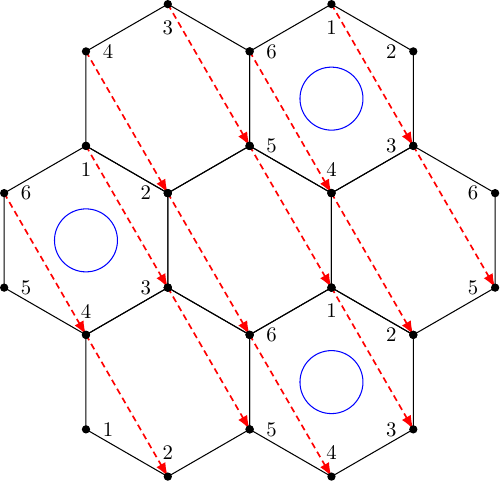}
    \caption{Schematic diagram for pseudo-translation $\mathcal{T}$ of vortex configurations. The red dashed arrows represent the translation vector, which also maps sublattice index $\gamma$ to $\gamma'$.}
    \label{fig:piover3jpg}
\end{figure}

We introduce an antiunitary transformation
$\mathcal{T}$:
\begin{equation}
\mathcal{T} a_{\mathbf{R},\gamma} \mathcal{T}^\dagger = a_{\mathbf{R}',\gamma'} ,
\end{equation}
The lattice index $\mathbf{R}'$ presents a uniform translation in real space shown in Fig. \ref{fig:piover3jpg}:
\begin{equation}
\mathbf{R}'=\mathbf{R}+\left(\sqrt{3}/2,-3/2\right) .
\end{equation} 

Consider 
$\mathcal{T}H(\alpha+\pi/3)\mathcal{T}^\dagger$. Since
\begin{equation}
\begin{aligned}
\mathcal{T} S_{\mathbf{R}, \gamma}^{a,b}(\phi_\gamma(\alpha+\pi/3)) 
\mathcal{T}^\dagger 
&=S_{\mathbf{R}',\gamma'}^{a,b}(\phi_{\gamma'}(\alpha)+\pi)\\
&=-S_{\mathbf{R}',\gamma'}^{a,b}(\phi_{\gamma'}(\alpha)),
\end{aligned}
\end{equation}
\begin{equation}
\begin{aligned}
\mathcal{T} S_{\mathbf{R}, \gamma}^{c}(\phi_\gamma(\alpha+\pi/3)) 
\mathcal{T}^\dagger 
&=-S_{\mathbf{R}',\gamma'}^{c}(\phi_{\gamma'}(\alpha)+\pi) \nonumber\\
&=-S_{\mathbf{R}',\gamma'}^{c}(\phi_{\gamma'}(\alpha)),
\end{aligned}
\end{equation}
we have
\begin{equation}
\mathcal{T} \mathbf{S}_{\mathbf{R},\gamma}(\alpha+\pi/3)
\mathcal{T}^\dagger
=-\mathbf{S}_{\mathbf{R}',\gamma'}(\alpha) .
\end{equation}
Because the translation $\mathbf{R}\rightarrow \mathbf{R}'$ 
preserves the $X, Y, Z$ bond of the Kitaev interaction,
\begin{equation} \label{pi/3conclusion}
\mathcal{T}H(\alpha+\pi/3,\theta,A_c)\mathcal{T}^\dagger=H(\alpha,\theta,A_c)
\end{equation}
By definition of $\mathcal{T}$,
\begin{equation}
\begin{aligned}
\mathcal{T} a_{\mathbf{k},\gamma}\mathcal{T}^\dagger 
&=\frac1{\sqrt{N}}\sum_{\mathbf{R}}
e^{i\mathbf{k}\cdot \mathbf{R}} 
a_{\mathbf{R'},\gamma'}\\
&=e^{-i\mathbf{k}\cdot (\sqrt{3}/2,-3/2)a} 
a_{-\mathbf{k},\gamma'} , 
\end{aligned}
\end{equation}
With this relation, \ref{pi/3conclusion} implies
\begin{equation}
\mathbf{P} \mathbb{H}^*(-\mathbf{k};\alpha+\pi/3) \mathbf{P}^T= \mathbb{H}(\mathbf{k};\alpha)  ,
\end{equation}
where $\mathbf{P}$ is the row-permutation matrix mapping index $\gamma$ to $\gamma'$, 
$\gamma+6$ to $\gamma'+6$. 
With $\mathbf{P}^T$ absorbed into $\mathbf{U}$, we see 
$\mathbb{D}(\mathbf{k};\alpha+\pi/3)=\mathbb{D}(-\mathbf{k};\alpha)$.

Next, we study $\alpha\rightarrow -\alpha$. By definition of $\phi_\gamma$,
\begin{equation}
\begin{aligned}
\phi_\gamma(-\alpha)=\begin{cases}
\phi_{\gamma'}(\alpha), &\text{vortex I} \\ 
\phi_{\gamma'}(\alpha)+\pi, &\text{vortex II}
\end{cases}
\end{aligned}
\end{equation}
with $\gamma'=\gamma+3$ mod 6. We introduce another transformation $\mathcal{P}$ 
which is unitary for vortex I and antiunitary for vortex II:
\begin{equation}
\mathcal{P} a_{\mathbf{R},\gamma} \mathcal{P}^\dagger = a_{-\mathbf{R},\gamma'} .
\end{equation}
Similar to the previous case, we can prove
\begin{align}
\mathcal{P} \mathbf{S}_{\mathbf{R},\gamma}(-\alpha)\mathcal{P}^\dagger&=\pm \mathbf{S}_{-\mathbf{R},\gamma'}(\alpha), \nonumber \\
\mathbf{P}'\mathbb{H}(-\mathbf{k};-\alpha) \mathbf{P}'^T&=\mathbb{H}(\mathbf{k};\alpha),\quad \text{vortex I} \\
\mathbf{P}'\mathbb{H}^*(\mathbf{k};-\alpha) \mathbf{P}'^T&=\mathbb{H}(\mathbf{k};\alpha),\quad \text{vortex II} \nonumber
\end{align}
where ``$+/-$'' corresponds to vortex I/II respectively and $\mathbf{P}'$ is a row-permutation matrix mapping $\gamma$ to $\gamma'$, $\gamma+6$ to $\gamma'+6$. This result implies $\mathbb{D}(\mathbf{k};-\alpha) = \mathbb{D}(\mp \mathbf{k};\alpha)$ for two vortex phases respectively. 

\section{Nearly degenerate vortex configurations}
The real-space vortex configurations presented in the main text are determined by the small local spin momentum obtained from DMRG simulations. In the vortex phase, we observed that the final vortex configuration in DMRG simulation exhibits weak dependence on the choice of initial wavefunction due to the nearly degenerate metastable states. To find out all the low-energy vortex configurations, we introduced a weak pinning field at one site near the boundary, namely adding the $-\vec{h}\cdot \vec{S}_{r_0}$ term to the Hamiltonian, where $\vec{h}$ is constrained in the [111] plane and $r_0$ labels the selected site at the boundary. In the first $5 \sim 10$ sweeps of the DMRG simulation, we apply a finite $|h|$ to align the spin at the $r_0$ site with the orientation of $\vec{h}$. Subsequently, we remove the pining field by setting $|h|=0$ and perform ten additional sweeps to ensure that the wave function converges to a low-lying state of the system. Depending on the orientation of $\vec{h}$, the final wave functions could possess nearly degenerate energy but distinct spin vortex configurations as discussed in the main text.

\end{document}